\documentclass[twoside,11pt]{article}
\usepackage{jmlr2e}

\usepackage{amsmath,amsfonts,amscd,amssymb,mathrsfs,subcaption}
\usepackage{graphicx,epstopdf,float,color} 
\usepackage{epsfig}
\usepackage{array,booktabs,paralist}

\usepackage[ruled]{algorithm2e}

\providecommand{\DontPrintSemicolon}{\dontprintsemicolon}
\renewcommand{\hat}{\widehat}

\newcommand{\no}{n_p}
\newcommand{\nv}{n_v}
\newcommand{\R}{R}
\newcommand{\D}{D}
\renewcommand{\citet}{\cite}

\begin{document}
\title{Phenotyping using Structured Collective Matrix Factorization of Multi--source EHR Data}
\author{\name Suriya Gunasekar \email{suriya@utexas.edu}\\	  
       \addr{University of Texas at Austin}\\
       {Austin, TX 78712, USA}      \AND
\name Joyce C. Ho \email{joyce.c.ho@emory.edu} \\
       \addr{Emory University}\\
       {Atlanta, GA 30322, USA}     \AND
\name Joydeep Ghosh \email{jghosh@utexas.edu}\\
       \addr{University of Texas at Austin}\\
       {Austin, TX 78712, USA} \AND
\name Stephanie Kreml\footnote{Dr. Kreml annotated the clinical relevance of the candidate phenotypes from   competing models.}
	   \email{stephanie@accordionhealth.com}\\
       \addr{Accordion Health, Inc.}\\
       {Austin, TX 78759, USA}\AND
\name Abel N Kho \email{a-kho@northwestern.edu}\\
       \addr{Northwestern University} \\
       {Chicago, IL 60611, USA}\AND
\name Joshua C Denny \email{josh.denny@vanderbilt.edu}\\
       \addr{Vanderbilt University} \\
       {Nashville, TN 37232, USA} \AND
\name Bradley A Malin \email{b.malin@vanderbilt.edu}\\
        \addr{Vanderbilt University} \\
       {Nashville, TN 37232, USA}\AND
\name Jimeng Sun \email{jsun@cc.gatech.edu}\\
       \addr{Georgia Institute of Tech.} \\
       {Atlanta, GA 30332, USA}       \AND
}
\maketitle

\begin{abstract}
The increased availability of electronic health records (EHRs) have spearheaded the initiative for precision  medicine using data driven approaches. Essential to this effort is the ability to identify patients with certain medical conditions of interest from simple queries on EHRs, or EHR-based phenotypes.  
Existing rule--based phenotyping approaches are extremely labor intensive. Instead, dimensionality reduction and latent factor estimation techniques from machine learning can be adapted for phenotype extraction with no (or minimal) human supervision.  

We propose to identify an easily interpretable latent space shared across various sources of EHR data as potential candidates for phenotypes. By incorporating multiple EHR data sources (e.g., diagnosis, medications, and lab reports) available in heterogeneous datatypes in a generalized  \textit{Collective Matrix Factorization (CMF)}, our methods can generate rich phenotypes.  Further, easy interpretability in phenotyping application requires sparse representations of the candidate phenotypes, for example each phenotype derived from patients' medication and diagnosis data should preferably be represented by handful of diagnosis and medications, ($5$--$10$ active components).   We propose a constrained formulation of CMF for estimating sparse phenotypes. We demonstrate the efficacy of our model through an extensive empirical study on EHR data from Vanderbilt University Medical Center. 
\end{abstract}


\section{Introduction}
Mining electronic health records (EHRs) can drastically improve clinical care and facilitate knowledge discovery.
In particular, EHR-driven phenotyping which involves identification of a set of clinical features or characteristic indicative of a medical condition from EHR data has been a key focus of EHR data analyses \citep{Collaboratory:Mr-x2zyZ}.
Phenotypes are important for targeting patients for screening tests and interventions, improving multisite clinical trials, and to support surveillance of infectious diseases and rare disease complications.
While existing efforts (e.g., eMerge Network, Phenotype KnowedgeBase, and the SHARPn program) have illustrated the promise of EHR--driven phenotypes, state of the art phenotype development generally requires an iterative and collaborative effort between clinicians and IT professionals to compose a series of rules for reproducible queries of EHR databases \citep{hripcsak2013next,Newton:2013fx}. A single phenotype takes substantial time, effort, and expert knowledge to develop.
Data mining tools such as support vector machines \citep{Carroll:2011ue}, active learning approaches \citep{Chen:2013es} and inductive logic programming \citep{Peissig:2014hp}, have been recently used to partially automate the phenotyping process.
Yet, these work require annotated samples to obtain good performance. As such annotations are expensive and time consuming to obtain, it is of interest to investigate unsupervised learning tools for automated phenotyping.

Phenotyping can be viewed as a form of dimensionality reduction of EHR data, where each phenotype or medical condition of interest represents  a latent space \citep{hripcsak2013next} and the rich literature in the field of machine learning for latent space estimation can be suitably adapted to automate and speed up the phenotype extraction process.
Several factors contribute to the quality of phenotypes extracted from EHR data, and it is advantageous to consider these factors in  choosing the appropriate dimensionality reduction tools for phenotyping. A review of the top 10 phenotypes across different studies showed that several data sources are typically used to define a phenotype \citep{shivade2014review}. Additionally, EHR data is commonly available in heterogeneous datatypes. For example, laboratory test results are often in the form of a real--valued number, patient demographic information can be encoded as a binary value, and procedure codes contain the number of times, a non-negative integer, the procedure is performed.  Thus, an automated phenotyping process that can incorporate data from heterogeneous datatypes and diverse sources can help identify rich existing as well as novel medical concepts.

Recent work has illustrated the promise of tensor factorization to generate phenotypes with minimal human supervision \citep{Ho2014:KDD, Wang:2015ip,henaoelectronic}
Latent space shared by various modes of higher order tensors are easier to interpret; and also more accurately capture the multi--source nature of phenotypes.
However, rich multi--way interactions required to form tensors is often not available in existing EHR data, for example, in a simple $3$rd order patient-diagnosis-medication tensor, the $(i,j,k)^\text{th}$ entry of the observation requires detailed  information on the number of times patient $i$ was prescribed medication $k$ in response to diagnosis $j$. In practice, much of the EHR data is available in flat formats that are more readily represented as matrices rather than tensors, e.g., a patient-diagnosis and a patient-medication matrix. Moreover, maintaining infrastructure to record and store higher order multi--way interactions is resource--intensive as the number of such possible interactions exponentially increase with each additional source. Alternatively, tensors constructed by approximating higher order interactions from flat format data could  lead to noisy correlations and biased results. 
These motivate the exploration of tools that directly work with multiple sources of matrix valued data.

In this paper, we propose unsupervised models for learning phenotypes from EHR data that are available as a collection of matrices. \textit{Collective Matrix Factorization (CMF)} \citep{sigo08} is an effective tool for identifying a latent space shared across multiple sources of data. In CMF, a collection of related matrices are jointly factorized into low--rank factors that are shared across the entire collection. For the phenotyping application, we introduce various structural and methodological modifications to the basic CMF model towards enhancing  interpretability of candidate phenotypes. 
\begin{asparaitem}



\item \textit{Heterogeneous datatypes:} Each source of EHR data can contain diverse datatype representations, such as numeric, count, or integer elements. Thus, it is desirable to use loss functions that are appropriate for the data in each  source.
We observe that a class of divergences called the \textit{Bregman divergences} are appropriate for our application as 
 this class includes divergence functions appropriate for various datatypes,  including continuous real--valued, binary and count data. 

\item \textit{Collective Factorization:} 
The challenge in effectively combining heterogeneous divergences in a collective matrix factorization is that such divergences often span different numerical scales and simple unweighted combinations tend to overfit datatypes or source matrices whose divergences are in the higher numerical range. We propose an effective heuristic approach to estimate appropriate weights for individual source matrices.

\item \textit{Non--negativity and Sparsity:} Physically interpretable latent factors are necessary to extract clinically meaningful phenotypes from EHR data.  Non--negative matrix factorization (NMF) \citep{paatero1994positive,lee1999learning} in comparison to the more traditional \textit{principal component analysis (PCA)} provides better interpretability of the low--rank factors as sum--of--parts representation. Such non--negativity constraints can be readily extended into the CMF framework. Further, sparsity of  latent factors representing the phenotypes plays a crucial role in the usefulness of the phenotypes as human experts need to analyze the factors and conduct further investigation to validate its clinical relevance. Thus, each phenotype should be ideally  be represented by very few active components ($\leq 10$ non--zero loading of entities) from each source. In one of the proposed variations of the generalize collective NMF formulation, convex sparsity inducing constraints are introduced to enhance the interpretability of extracted latent factors. 
\end{asparaitem}

We empirically assess the proposed model on real EHR data from Vanderbilt University. The clinical relevance of the extracted phenotypes are evaluated by domain experts. 
\section{Related Work} \label{sec:relatedwork}
In this paper, we address phenotype extraction from EHR data using unsupervised dimensionality reduction techniques. Inferring low--dimensional representation of matrix data is a fundamental problem in machine learning and a complete review of related techniques is beyond the scope of the paper. We briefly discuss just the key models that are most relevant to the work in this paper. PCA \citep{fodor2002survey,jolliffe2002principal}, the most popular and widely used tool dimensionality reduction, learns latent factors as low rank  matrices whose values are unconstrained and can contain both positive and negative entries. However, in many applications it is desirable to interpret the low rank factors as physical concepts and negative entries often contradict physical reality. This motivated a related line of dimensionality reduction techniques called the Non--Negative Matrix factorization (NMF) \citep{paatero1994positive,lee1999learning}.
Several existing work extend matrix factorization tools to analyze data from multiple matrices. Collective matrix factorization (CMF) and its non--negative variants \citep{sigo08} incorporate information from multiple sources of matrix data using shared latent variables/factors. Alternatively, regularized NMF variants have been proposed combining data from multiple sources \citet{zhang2011novel,liu2013multi}. The tools for matrix valued data have also been generalized to higher order tensors, or multi--way arrays (see \citep{kolda2009tensor} for a review). 
Variants of non-negative tensor factorization (NTF) based on CANDECOMP--PARAFAC, one of the most popular tensor decomposition models, and its applications to extract interpretable latent/hidden factors have been proposed, e.g. \cite{Cichocki:2009vo,Lee:2007tv,Cichocki:2008vh, Acar:2011vx,yilmaz2011generalised,Acar:2014ft} and references therein.
However, most of these methods primarily utilize the least square loss and may not be appropriate for all data types.
We build on these existing work and propose techniques for efficiently extracting structured latent factors from multiple sources of heterogeneous EHR data. The primary focus is on the interpretability of the  low--dimensional factors as meaningful phenotypes. 

Although existing phenotyping methods rely on a labor--intensive supervision, unsupervised models have been proposed to leverage the vast amount of EHR data for automatic phenotype discovery. These models include the use of probabilistic graphical models to cluster patient's longitudinal trajectories \citep{schulam2015clustering}, deep learning to detect characteristic patterns in clinical time series data \citep{Che:2015cy}, and generative models on static data \citep{chen2016identifying}. Yet these methods are not scalable and are ill-suited for incorporating data from patients over a prolonged period of time ($6+$ months).  Recent work has illustrated the promise of NTF to generate phenotypes with minimal human supervision using data over several years \citep{Ho2014:JBI, Ho2014:KDD, Wang:2015ip, henaoelectronic}. 
However, as noted earlier, a tensor representation is not always available in EHR data, at least not without introducing assumptions and potentially biasing the results.


\section{Phenotyping from EHR Data}
The notations used in the rest of the paper are summarized in Table \ref{tab:notation}.  We assume that the patient EHR data from $V$ sources, such as medications, diagnosis, laboratory measurements, etc. are available as  matrix valued data whose rows correspond to a common set of  patients, and columns represent entities from the respective sources (medications, diagnosis, laboratory measurements, etc.). 
Let $\no$ denote the number of patients, and for each source $v\in\{1,2,\ldots,V\}$, let $\nv$ denote the number of unique entities within the source $v$. 
The collection of $V$ matrices containing EHR data from multiple sources is denoted by $\mathcal{X}=[X_v]_{v=1}^{V}$, where $X_v\in\mathbb{R}^{\no\times\nv}$ denotes the matrix data from source $v$. 

\begin{table}

\begin{center}
\begin{tabular}{l l}
\toprule
\textbf{Notation} & \textbf{Description} \\
\midrule
$[N]$ & Set of integers $[N]=\{1,2,\ldots,N\}$\\
$X^{(k)}$ & Column $k$ of a matrix $X\in\mathbb{R}^{n_r \times n_c}$\\
\textbf{Input}&\\
$v\!=\!1,2,\ldots,V$ & Index over $V$ sources of EHR data, e.g. medication, diagnosis, etc.\\
$\no$ & Number of patients\\
$\nv$ & Number of entities in source type $v$\\
$X_v\in\mathbb{R}^{\no\times \nv}$ & EHR data matrix from source $v$\\

$\mathcal{X}=[X_v]_{v=1}^V$ & Collection of $V$ EHR data matrices\\
$\D_v$ & Bregman divergence appropriate for approximating $X_v$\\

\textbf{Estimates}&\\
$\widehat{\mathcal{X}}=[\widehat{X}_v]_{v=1}^V$ & Estimate of $\mathcal{X}$  from models\\

$W\in\mathbb{R}^{\no\times \R}$& Patients' loading along 
the $R$ dimensional latent space of interest\\
$H_v\in\mathbb{R}^{\nv\times \R}$ & Latent factor representation for features in source $v$\\
$b_{v}\in\mathbb{R}^{\nv}$ & Bias factors associated with columns of the data matrix $X_v$\\

\bottomrule
\end{tabular}
\end{center}
\caption{Summary of Notations \label{tab:notation}}
\end{table}
\subsection{Dataset Overview}\label{sec:data}
We evaluate our models on an EHR data set from Vanderbilt University Medical Center. This section contains a brief exploration of the data and the empirical results.

We used de-identified electronic medical records corresponding to the first $\sim$10,000 patients in BioVU\footnote{https://victr.vanderbilt.edu/pub/biovu/}, the Vanderbilt DNA databank, spanning over 20 years. The details of the inclusion and exclusion criteria for the databank are described in \citet{Ritchie:2010jd}. For evaluation purposes, we focused on the case and control patients for type--2 diabetes and resistant hypertension. These patients and their labels were selected by using the respective rule--based phenotype algorithms defined in the Phenotype KnowledgeBase\footnote{https://phekb.org}. We emphasize that  labels from these rule-based algorithms were \textit{not} used in our phenotyping models which are learned in a completely unsupervised setting. 

\begin{figure*}[htb]
\centering
\includegraphics[width=0.8\textwidth]{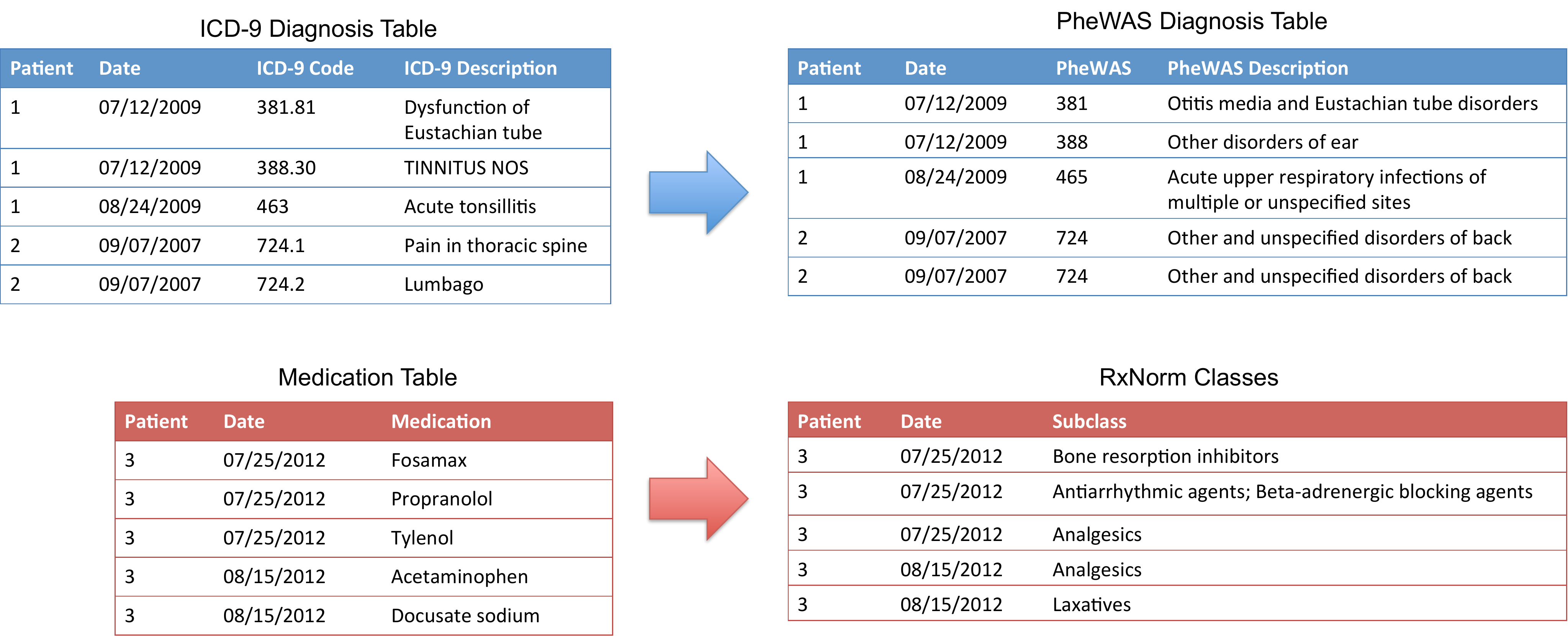}
\caption{Examples of the aggregation from ICD-9 diagnosis codes to PheWAS code groups and original medications to the MeSH pharmacological actions classes.}
\label{fig:mapping}
\end{figure*}

Although our model is general enough to be applied to multiple data types, we work with counts of diagnoses and medications for evaluation purposes. The diagnosis codes, in the form of International Classification of Disease, 9th edition (ICD-9) codes, were grouped using the PheWAS code groups\footnote{http://phewas.mc.vanderbilt.edu/}, a custom-developed hierarchy which currently contains $\sim 1600$ groups. Medications were aggregated based on Medical Subject Headings (MeSH) pharmacological actions provided by the RxClass REST API, a product of the US National Library of Medicine. Note that a medication may belong to multiple categories. Figure \ref{fig:mapping} provides example aggregations performed on the original table for the purpose of our study. 

Finally, BioVU dataset assigns an index (reference) date to each patient, which corresponds either to the date where the criteria was met (case patients) or the last encounter date (control patients). The EHR records of patients falling in the date range of one year prior to their index date up until the index date were used in our experiments. Any patient without at least one diagnosis and medication during the relevant time period was not included in our study. Our resulting data set contains 2039 patients, 936 diagnosis groups, and 161 medication classes.

The dataset is summarized in Table~\ref{tab:data} and the top five diagnosis and medication categories that appear in our data are shown in Table \ref{tab:top-5}.

\begin{table}[htb]
\centering
\begin{tabular}{l l c c}
\toprule
$v$ &
Source Matrix $X_v$ &  $\no\times n_v$ &  Datatype \\
\midrule
$1$ & Patient--Diagnosis & $2039\times 936$ &  Count \\
$2$ & Patient--Medication & $2039\times 161$ & Count \\
\bottomrule
\end{tabular}
\caption{Dataset summary.
\label{tab:data}}
\end{table}
\begin{table}[htb]
\centering
\begin{tabular}{l p{12cm}}
\toprule 
Source & Top five entities\\
\midrule 
Diagnosis & Hypertension; Incision, excision, and division of other bones; Ischemic Heart Disease; Secondary malignant neoplasm of respiratory and digestive systems; Disorders of lipoid metabolism.\\
Medication & Analgesics; Vitamins; Anticonvulsants; Anxiolytics, sedatives, and hypnotics; Antihyperlipidemic agents.\\
\bottomrule
\end{tabular}
\caption{The top five diagnosis and medications of the patients in our study.
\label{tab:top-5}}
\end{table}



\section{Structured Collective Matrix Factorization for Phenotyping}\label{sec:model} 
For each source $v\in[V]$, we approximate $X_v$ by structured estimates $\hat{X}_v$ which incorporates model constraints appropriate for effective phenotyping. 

\subsection{Heterogeneous Datatypes}\label{sec:divergence} In EHR data from multiple sources, each source matrix $X_v$ may contain data represented in diverse datatypes (e.g., binary values for demographics, count values for medications, or continuous values for laboratory measurements). In our phenotyping models, the data fidelity of $\widehat{X}_v$ is quantified using an appropriately chosen source--specific divergence $D_v(X_v,\widehat{X}_v)$. The divergence functions are selected from a class of \textit{Bregman divergence} defined as follows:
\begin{definition}[Bregman Divergence]\label{def:bd}\normalfont
 Let $\phi$ be a strictly
convex function differentiable in the relative interior of $\text{dom}(\phi)$. The \textit{Bregman divergence} (associated with $\phi$) between $x\in\text{dom}(\phi)$ and $y\in\text{ri}(\text{dom}(\phi))$ is defined as:\[D^\phi(x,y)=\phi(x)-\phi(y)-\langle \nabla \phi(y),x-y\rangle.\]
\end{definition} 
The motivation for using Bregman divergences are two fold. 
Bregman divergences include rich classes of loss functions that are appropriate for a variety of datatypes including (weighted) squared loss  for continuous valued data, logistic loss for binary valued data, and  generalized KL divergence for count valued data among others \citep{banerjee2005clustering}. These loss functions  are also equivalent to the negative log--likelihood of members of exponential family distributions including Gaussian, Bernoulli, Poisson, exponential among others \citep{banerjee2005clustering}. Thus, the domain knowledge of data distribution can be potentially incorporated in choosing the appropriate divergence. 
Secondly, Bregman divergences are strictly convex and differentiable in the first parameter, and accurate and tractable estimators for $\hat{X}_v$ can be developed using gradient descent and alternating minimization algorithms. 


In our dataset described in Section~\ref{sec:data}, as both the matrices described in Section~\ref{sec:data} have count valued data we use the generalized KL divergence given by the following equation as the divergence for both sources:
\begin{equation} D(X,\hat{X})=\sum_{ij} \hat{X}_{ij}-X_{ij}+X_{ij}\log{\frac{X_{ij}}{\hat{X}_{ij}}}.
\label{eq:genKL}
\end{equation}

\subsection{Generalized Collective NMF (CNMF)}\label{sec:cnmf}
As noted earlier, non--negativity constraints on the patient loading and latent factor matrices allow for better interpretability as sum--of--parts representation.  We propose a generalized collective NMF (CNMF) as a basic model for extracting phenotypes from  multiple sources of patient data available in heterogeneous datatypes. In Section~\ref{sec:sicnmf}, we introduce additional structures to enhance interpretability. 

Each source of EHR data $v$ is associated with a structured latent factor  matrix $H_v\in\mathbb{R}^{\nv\times R}$, and these factors jointly span a shared latent space. The columns of $H_v$ concatenated across the $V$ sources are potential candidates for phenotypes. The loading of the patients along these latent dimensions are given by the matrix $W\in\mathbb{R}^{\no\times R}$. Additionally, the raw EHR data often contains generic features  that are not necessarily indicative of any medical condition of interest. For example, medications like pain reliever, laboratory measurements like body temperature, etc. are frequently encountered in patient data, but are not discriminative of patient conditions. EHR data from such frequent and non--discriminative features are captured through an  explicit (and potentially dense) column or feature bias factor $b_{v}\in\mathbb{R}^{n_v}$ for each source $v$.

For $v\in[V]$, the source matrix $X_v$ is approximated as $WH_v^\top+\mathbf{1}b_{v}^\top$, where $\mathbf{1}$ is a vector of all ones in appropriate dimensions. We use a Bergman divergence $D_v$ appropriate for each source to measure the data fidelity of the estimate to the observed data. Finally, as the heterogeneous divergences are in different scales, we weight the divergence corresponding to each source using weight parameter $\alpha_v,v=1,2,\ldots,V$. The basic CNMF  estimator is given by the following optimization problem. 
\begin{equation}
\begin{aligned}
\hat{\mathcal{X}}=\;\underset{\{\hat{X}_v\}_{v\in[V]}}{\text{argmin }} &\sum_{v=1}^V\alpha_v\D_v(X_v,\hat{X}_v),\\
\text{s.t. }&\;\hat{X}_v=W H_{v}^\top+\mathbf{1}b_{v}^\top \text{ for } v\!=\!1,2,\ldots,V,\\
&\;W\in\mathbb{R}^{\no\times R}_+, \; H_v\in\mathbb{R}_+^{\nv\times R}, \; b_{v}\in\mathbb{R}_+^{\nv}.
\end{aligned}
\label{eq:cnmf}
\end{equation}

\subsubsection{Computing $\{\alpha_v:v=1,2,\ldots,V\}$}\label{sec:alpha}
As noted earlier, since the divergences associated with difference datatypes span different numerical scales, unweighted objective in \eqref{eq:cnmf} will tend to overfit the  matrices whose divergences are in the higher numerical range. We propose an effective heuristic approach to estimate contribution of each source matrix $X_v$ in the joint estimation. To motivate the idea, consider a source matrix $X_v$. If a joint factorization is not required, i.e. $W$ need not be shared, then the optimization problem in \eqref{eq:cnmf} can be solved as $V$ independent structured factorization $\widetilde{X}_v^\text{ind}=W_vH_v^\top+\mathbf{1}b_{v}^\top$ without the weights $\alpha_v$. In a preprocessing step, for each source and independent factorization of the form $\widetilde{X}_v^\text{ind}$ is learned by minimizing $D_v(X_v,\widetilde{X}_v^\text{ind})$ assuming the sources to be independent of each other. The resultant divergence from independent factorization is treated as the effective scale of divergence for each source. To assign equal importance to all source matrices, we choose $\forall v,\;\alpha_v=\frac{1}{D_v(X_v,\widetilde{X}_v^\text{ind})}$.

\subsection{Sparsity--inducing CNMF  (SiCNMF)} \label{sec:sicnmf}
As phenotypes learned form data analysis tools are further investigated by human experts, it is desirable that candidate phenotypes learned from EHRs are sparse combinations of the source entities, i.e., columns $H_v$ are sparse. 

To illustrate our  sparsity--inducing constraints for enhanced interpretability, we first consider a single source of EHR data matrix $X\in\mathbb{R}^{\no\times n}$ and an appropriate divergence function $D(.)$. 
 Explicit sparsity constraints on the factor matrix $H$ lead to  intractable combinatorial optimization problems. A commonly used convex surrogate for sparsity involves restricting the $\ell_1$ norm of the columns of $H$, i.e., constraints of the form $\{\|H^{(k)}\|_1\le s:  k\in[R]\}$, for some parameter $s$.  However, in \eqref{eq:cnmf} if the scaling  of  $W$ is unrestricted, then due to  multiplicative nature of the factorization, restrictions on norm of $H$ tend to be  ineffective as any scaling of $H$ can be easily absorbed by $W$. Thus, we additionally constrain the scale of $W$ using a Frobenius norm constraint of the form $\|W\|_F\le \eta$, for another parameter $\eta$. We note that,  $s$ and $\eta$ effectively work as single parameter due to the multiplicative update. Thus, without loss of generality, we fix $s=1$ and use  $\eta$ as a tunable parameter to control the sparsity level.

We propose the following generalized SiCNMF model as an extension of vanilla CMF which incorporates (a)  sparsity--inducing  and non--negativity constraints for enhanced interpretability,  (b) feature specific bias factors $\{b_{v}: v\in[V]\}$ to capture data specific offsets, and  (c) appropriately weighted heterogeneous divergences to handle varied datatypes.
\begin{equation}
\begin{aligned}
\hat{\mathcal{X}}=\;\underset{\{\hat{X}_v\}_{v\in[V]}}{\text{argmin }} &\sum_{v=1}^V\alpha_v\D_v(X_v,\hat{X}_v),\\
\text{s.t. }&\;\hat{X}_v=W H_{v}^\top+\mathbf{1}b_{v}^\top \text{ for } v\!=\!1,2,\ldots,V,\\
&\;W\in\mathbb{R}^{\no\times R}_+, \; H_v\in\mathbb{R}_+^{\nv\times R}, \; b_{v}\in\mathbb{R}_+^{\nv}, \\
&\; \|W\|_F\le \eta,  \|H_{v}^{(k)}\|_1=1\;\forall  k\in[R],
\end{aligned}
\label{eq:sicnmf}
\end{equation}
where recall that $H_{v}^{(k)}$ is the $k^\text{th}$  column of $H_v$, and $\alpha_v$ are either (a) all ones (unweighted SiCNMF), or (b) computed using the methodology described in Section~\ref{sec:alpha} (weighted SiCNMF). Note that the higher the value of $\eta$, the weaker the sparsity constraint. In the limiting case of $\eta=\infty$, the model is equivalent to the heterogeneous collective non--negative matrix factorization (CNMF) as scaling constraints of $H_v$ are captured by $W$.
\section{S\textsc{i}CNMF: Algorithm Details} 
For any set of Bregman divergences $\{D_v:v=1,2,\ldots V\}$ and positive parameters $\eta,\{\alpha_v\}>0$ , the optimization problem \eqref{eq:sicnmf} is convex in $[(H_v,b_{v})\; \forall v]$ when $W$ is fixed and vice versa. Our algorithm uses  alternating minimization to solve \eqref{eq:sicnmf} where each iteration alternatively minimizes $[(H_v,b_{v})\; \forall v]$ and $W$,  while keeping the other fixed. Each such component update involves minimizing a smooth convex objective subject to convex constraint set and is solved using projected gradient decent algorithm with backtracking line search to determine step size \citep{lin2007projected}. 


Recent work has shown that projected gradient methods are computationally competitive and have better convergence properties than standard multiplicative update approaches \citep{lin2007projected}. Moreover, compared to multiplicative updates, projected gradient descent based algorithms can be easily extended for convex constraints beyond simple non--negativity.
Although \citet{lin2007projected} ignore the KL divergence
problem as ill-defined, a more recent work \citep{Chi:2012jj}	 provide convergence for related tensor factorization task by  showing that the
convex hull of the level sets of the KL divergence problem is compact.
To project onto the simplex, we used the simple and fast algorithm proposed by Chen and Ye \citep{Chen:2011wz}.

Our algorithm to solve \eqref{eq:sicnmf} is described in Algorithm \ref{alg:projected}.


\begin{algorithm} [htb] {
\DontPrintSemicolon
\SetKwInOut{Input}{Input}\SetKwInOut{Output}{Output}
\Input{EHR data $X_v$, $D_v(.)$ for $v=1,2,\ldots,V$}
\textbf{Parameters} :{ Divergence weights $\{\alpha_v\}$ and tunable  sparsity inducing parameter $\eta\in(0,\infty)$}

\Output{Patient loadings $\hat{W}$, factors/phenotypes $\{\hat{H}_v\}$ and feature  biases $\{\hat{b}_{v}\}$}

\While{\text{not converged}} {
	$\hat{W}=\begin{array}{l} \underset{W\ge 0}{\text{argmin}}\;\; \displaystyle{\sum_{v=1}^V}\alpha_vD_v(X_v,W\hat{H}_v^\top+ \mathbf{1}\hat{b}_{v}^\top)\\ \text{s.t.  }\quad\quad |W\|_F\le \eta,\; W\ge 0.\end{array}$

	\For {$v\in[V]$} {	
$\hat{H}_{v},\hat{b}_{v}=\begin{array}{l} \underset{H_v\ge0,b_{v}\ge0}{\text{argmin}}\;\; D_v(X_v,\hat{W}H_v^\top+\mathbf{1}b_{v}^\top)\\
\text{s.t. } \quad\quad\quad\;\,\forall k, \|H_v^{(k)}\|_1=1.\end{array}$	
	}
}
}
\caption{Alternating minimization for \eqref{eq:sicnmf} using projected gradient descent}\label{alg:projected}
\end{algorithm}


\section{Experiments}
The generalized KL--divergence \eqref{eq:genKL} is used as loss function for both matrices (patient by diagnosis and patient by medication) in the collective matrix factorization models as well as the baselines described in the following subsection.

\subsection{Baseline Models}\label{sec:baseline}
The primary focus of the paper is the clinical relevance of candidate phenotypes obtained from unsupervised dimensionality reduction techniques.  Since the Vanderbilt data contains flat files associated with the diagnosis codes and medications, construction of the patient--medication and patient--diagnosis matrices for the collective matrix factorization models were straightforward. Our models of CNMF \eqref{eq:cnmf} and SiCNMF \eqref{eq:sicnmf} are compared with two baseline models described below:
\begin{itemize}
\item {\bf Non--negative matrix factorization (NMF) \citep{lee1999learning}}: In order to evaluate traditional NMF in identifying a shared latent space, we aggregate the patient information into a third matrix, diagnosis by medication, wherein each element represents the number of patients who have at least one occurrence of both the diagnosis and the medication during our one year time window. It is important to note that under this construction, a patient with two encounters almost one year apart, one with the diagnosis A and one with medication B would be counted in the $(A,B)^{\text{th}}$ entry of the matrix. A non--negative matrix factorization model with the generalized KL divergence as the objective and no sparsity constraints is performed on the diagnosis by medication matrix.
\item {\bf Marble \citep{Ho2014:KDD}}: Marble is a sparse non--negative tensor factorization model that has been used to obtain highly effective and interpretable phenotypes provided a multiway tensor EHR data is available. However, our dataset does not have rich multi--way interactions to easily construct tensors. For example, in a patient--diagnosis--medication tensor, a entry $x_{ijk}$ denotes the number of times a patient $i$ was prescribed medication $k$ in order to treat a diagnosis $j$. To construct tensors from available flat files,  we approximate these interactions  by assuming that a medication was used to treat a specific diagnosis if both diagnosis and medication occur within a one week time interval, that is the counter for $x_{ijk}$ is incremented if patient $i$ was prescribed medication $k$ within one week of an encounter with diagnosis $j$. Marble applied to this approximated tensor is our second baseline.
\end{itemize}

The baselines described above are compared to three CMF based models described in this paper: (a)  CNMF \eqref{eq:cnmf} which does not incorporate the sparsity inducing constraints, (b) unweighted SiCNMF  which incorporates sparsity--inducing constraints proposed in Section~\ref{sec:sicnmf}, but uses a simple aggregation of various source divergences, i.e., solves \eqref{eq:sicnmf} with $\alpha_v=1$ for all $v\in[V]$, and finally (c) weighted SiCNMF which incorporates both the sparsity--inducing structure and the weights $\alpha_v$ computed using the heuristic described  in Section~\ref{sec:alpha}. 

All the models described above involve non--convex optimization and the estimates from the algorithm are  sensitive to initialization. To mitigate this issue from local minima, we run each algorithm independently multiple times and pick the run with best fit to the objective. All the competing models learn a  $R=20$ rank factorization.

\subsection{Sparsity--accuracy trade off: Data fit}

The sparsity of the candidate phenotypes plays a crucial role in the interpretability and wider applicability of the estimates. Concise representations allow domain experts to more easily reason about a particular group of patients queried using the phenotypes. As noted earlier, while non--negativity constraint in matrix and tensor factorization inherently induce sparsity as a by--product, there is no explicit control over the sparsity levels. Thus in order to deriving extremely sparse phenotypes involve, we introduce a sparsity inducing regularization, whose sparsity levels can be controlled by an tunable knob of $\eta$ in \eqref{eq:sicnmf}.

\begin{figure*}[htb]
\centering
\includegraphics[width=\textwidth]{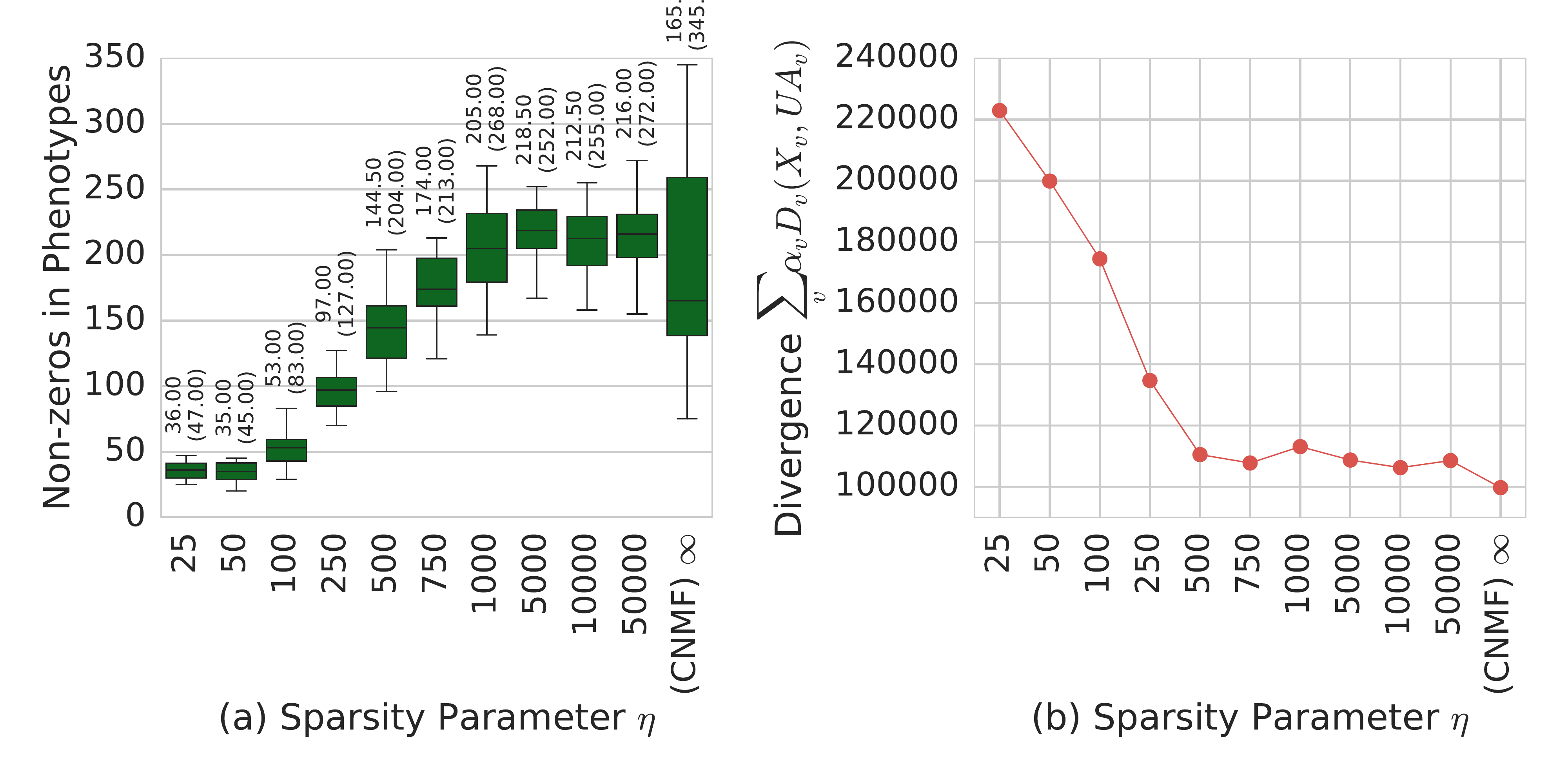}
\caption{Sparsity--accuracy trade-off in data fit of weighted SiCNMF. Sparsity  is measured as the  median number of non-zero entries in columns of the phenotype matrices concatenated from all sources $\{\hat{H}_v:v=1,2,\ldots, V\}$. (a) Each box plot represents the spread of the  number of non--zeros  in  $R=20$ candidate phenotypes learned from weighted SiCNMF  using $\eta$ represented along the x--axis in \eqref{eq:sicnmf}.   (b) Plot of decay of divergence between the fitted estimate and the observed data as the sparsity constraint is relaxed using higher $\eta$. Note that the values of $\eta$ along x--axis are not in linear scale and higher values correspond to weaker sparsity--inducing regularization.
\label{fig:sparsity1}}
\end{figure*}

The expected sparsity--accuracy trade-off in the data fit can be observed in Figure~\ref{fig:sparsity1}.  Note that higher values of $\eta$ in  \eqref{eq:sicnmf} correspond to a weaker sparsity constraints as the $W$ factor can more easily absorb the scaling constraint on $H_v$. 

\subsection{Type-2 diabetes and Resistant hypertension prediction}
\begin{figure*}[htb]
\centering
\includegraphics[width=0.8\textwidth]{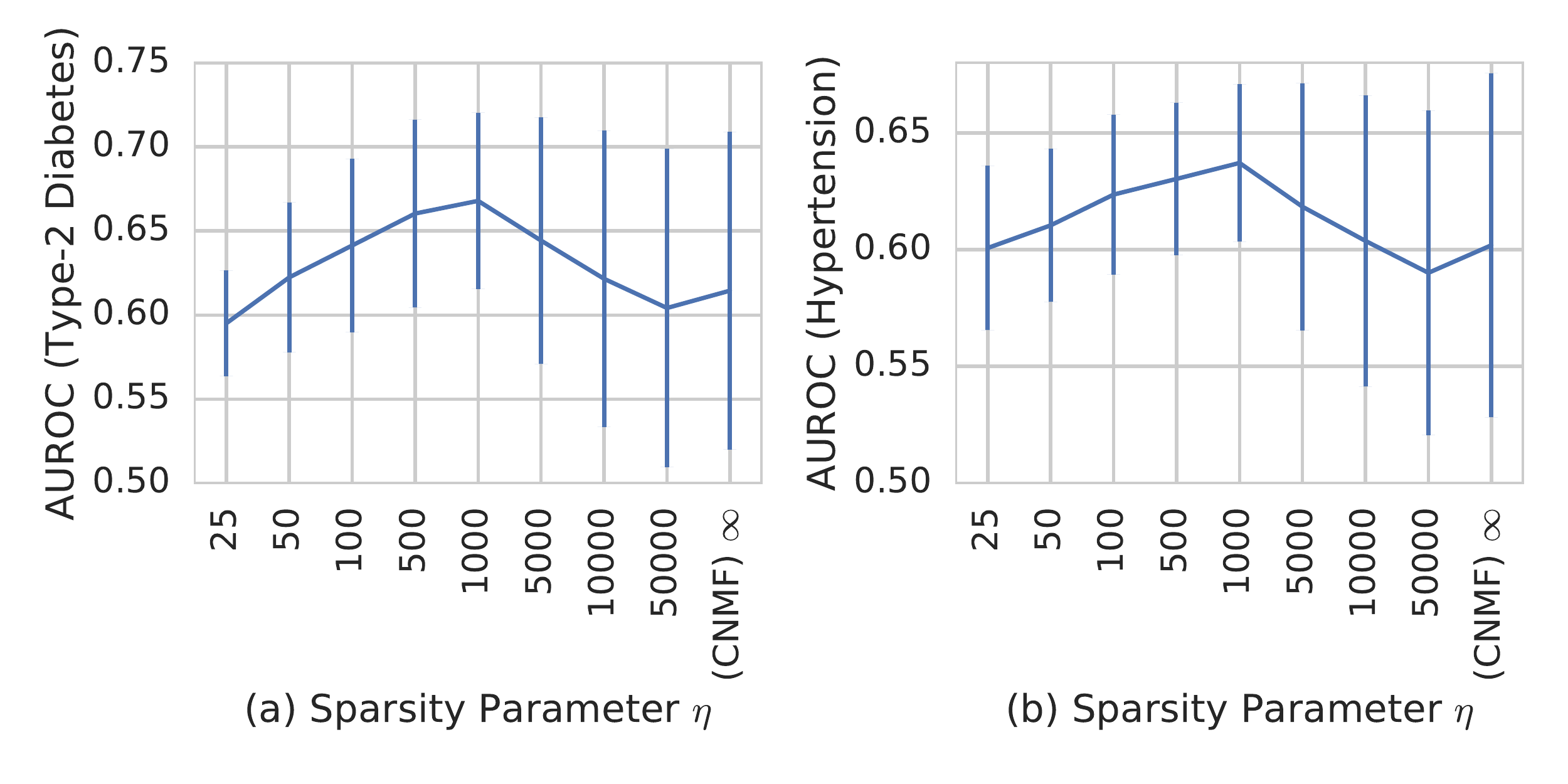}
\caption{Sparsity--accuracy tradeoff in prediction of (a) Type--2 diabetes and (b) resistant hypertension. The results are for weighted SiCNMF, but similar tradeoff was also observed for unweighted SiCNMF. Note that x--axis is not linear and higher $\eta$ leads to lower sparsity (more number of non-zeros in phenotype representations)}\label{fig:pred-sparsity}
\end{figure*}

With relaxed sparsity constraints, while a monotonic decay of objective function on training data fit  as observed in Figure  \ref{fig:sparsity1} is expected, such a monotonic accuracy trade-off does extend for predictions on held out test datasets. Besides improving interpretability, the sparsity constraints further function as regularization to prevent overfitting.  

To quantitatively evaluate the effectiveness of the extracted phenotypes, we consider the classification problem of predicting two chronic conditions prevalent in the patient population of our dataset (Section~\ref{sec:data}): (a) type-2 diabetes, and (b) resistant hypertension. As described in Section~\ref{sec:data}, for each patient in our dataset, the class labels for these chronic conditions were estimated from rule-based phenotyping algorithm from PheKb. 

Our full dataset of $\sim2000$ patients is divided into $5$ stratified cross validation folds  of $80\%$ training and $20\%$ test patients.  For each cross-validation fold, the models described in Section \ref{sec:baseline} were applied on training EHR dataset to extract the phenotype matrices $\{\hat{H}_v^\text{train}:v\in[V]\}$. We clarify that,  for all the competing models, the phenotypes (latent factors) were extracted (a) only from EHR data of patients in the  training set, and  (b) the estimates were learned in a completely unsupervised setting. In particular, the test EHR data and the labels were \textit{not} used in  the phenotype extraction phase. 

For each patient, the $R$ dimensional loading along the phenotype/latent space spanned by $\{\hat{H}_v^\text{train}:v\in[V]\}$ is used as features for learning the classifiers. Such representations are  computed by projecting the EHR matrix into the fixed phenotype factors. For CMF variants,  the features for a patient with EHR $[X_v^\text{patient}]$ is given by: $W^\text{patient}=\underset{W\ge 0}{\text{argmin}} \sum_v\alpha_v{D}_v(X_v^\text{patient}, WH_v^\text{train}+\mathbf{1}b_{v}^{\text{train}\top}).$

The sparsity--accuracy trade-off in prediction performance on held out dataset is plotted in Figure \ref{fig:pred-sparsity}. Although, the predictive performance at various $\eta$ levels are statistically comparable, the mild regularization effect of sparsity constraints can be observed the plots.

\subsection{Sparsity and Prediction Comparison to Baseline Models}
So far in the experiments, we have mainly focused on the effect of sparsity parameter on data fit and prediction performance. In this subsection, we compare the performance CMF based estimators to strong baselines models. 
\subsubsection{Sparsity}
\begin{figure}[htb]
\centering
\includegraphics[width=\columnwidth]{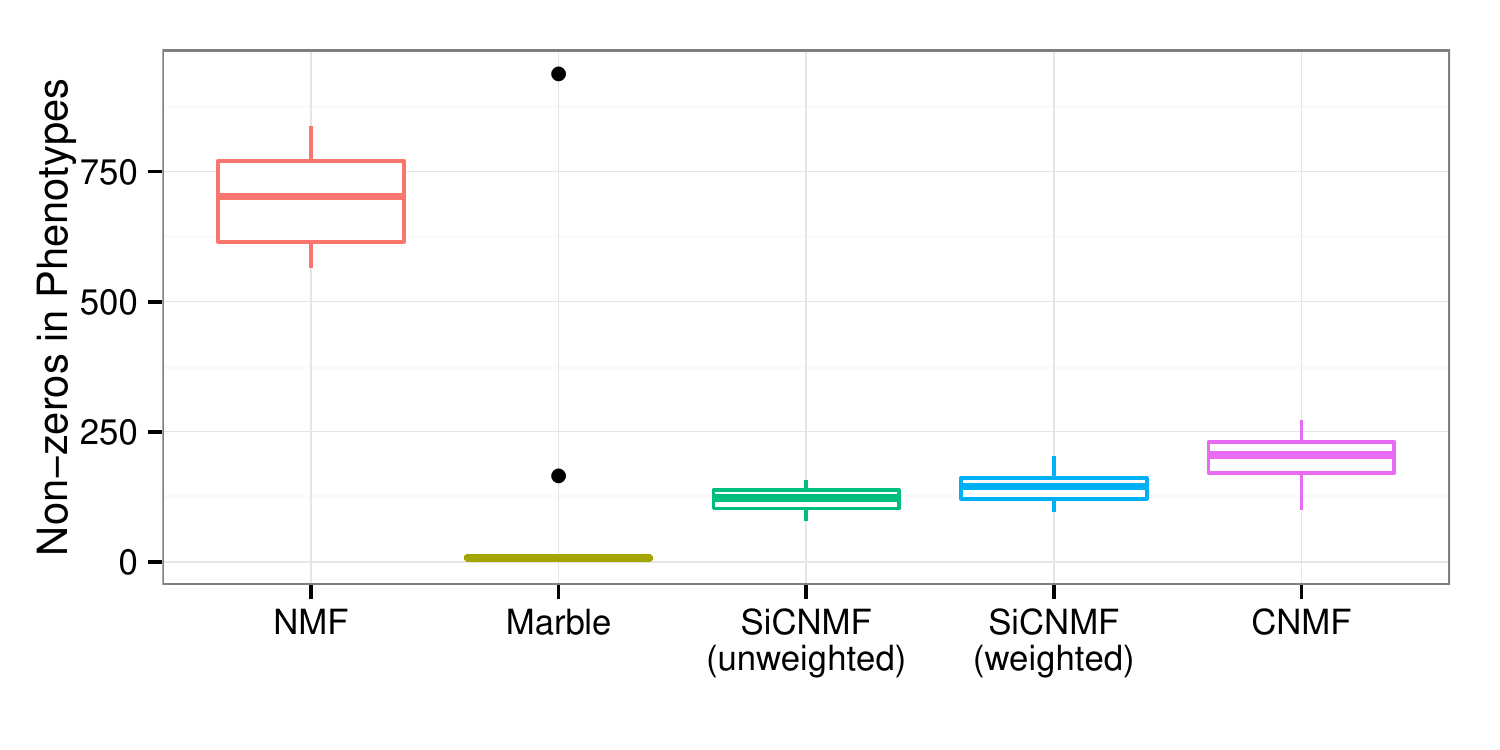}
\caption{Box plots showing the inherent sparsity induced by the models.}
\label{fig:sparsity}
\end{figure}

The sparsity patterns obtained by the competing phenotyping algorithms described in Section~\ref{sec:baseline} are compared in Figure \ref{fig:sparsity}. As expected,  the sparsity of SiCNMF models are better than those of non--sparsity--inducing  CNMF and NMF models. NMF \citep{lee1999learning} on dense aggregated data which does not incorporate  explicit sparsity constraints learns dense factor matrices. We note that CNMF  models multiple sparse matrices jointly learns much sparser factors compared to NMF on single aggregated matrix. 
Marble \citep{Ho2014:KDD} induces sparsity by truncation and achieves the best sparsity performance. 

\begin{figure*}[htb]
\centering
\includegraphics[width=\textwidth]{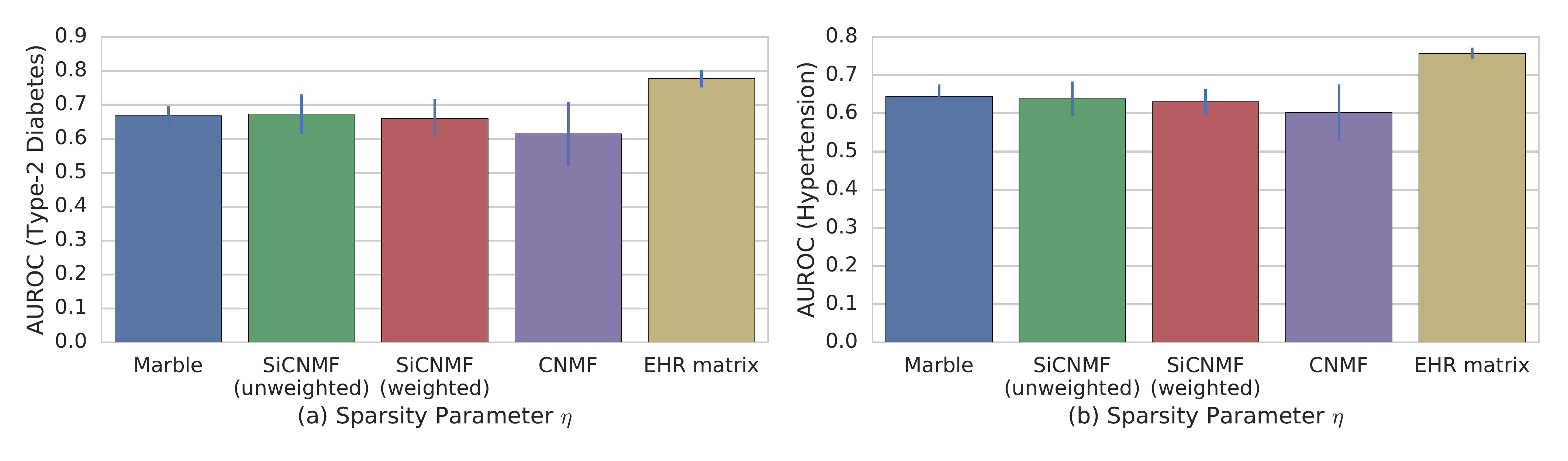}
\caption{Accuracy—sparsity tradeoff in prediction\label{fig:pred-models}}
\end{figure*}
\subsubsection{Prediction}
The classification performance of baseline models for predicting type-2 diabetes and resistant hypertension are compared in Figure~\ref{fig:pred-models}. We note here that as NMF uses aggregated data of patients and there is no easy approach to learn individual patient representations in the phenotype space. Thus, we exclude NMF from this set of experiments.
 Instead we use the classifiers learned on \textit{full} concatenated EHR matrix as an additional baseline for prediction performance. Note that the concatenated EHR matrix has $> 1000$ features compared to the $20$ dimensional representation of the rest of the models. It is observed that the phenotype based models with $20$ dimensional feature representation have comparable performance. However, the classifiers from full EHR matrix with $>1000$ features outperforms the phenotype--based models. While the EHR matrix provides a richer set of features for prediction performance, the high dimensional EHR data are not not useful for phenotyping applications and interpretability.

\subsection{Clinical Relevance of  Phenotypes} \label{sec:relevance}
The  phenotypes extracted from the models described above are evaluated by a human expert for clinical relevance. For reasonable evaluation of the phenotypes by humans,  it is desirable that each phenotype be represented by a very small number of  diagnoses and medications groups. Based on a round of feedback from our clinical experts, in post processing, we retain just the top $5$ medications and top $5$ diagnosis from phenotypes learned from \textit{all} the models for evaluation in this section.

We evaluated the clinical relevance of the resulting phenotypes from the phenotyping models by conducting a survey with a domain expert. The domain expert was given $20$ phenotypes from each model to assess and were not informed apriori the correspondence between the models and the results. For each of the individual phenotypes, the experts assigned one of three values: \textbf{(1) yes -- it was clinically meaningful, (2) possible -- the phenotype has some clinical meaningfulness, and (3) no -- it was not meaningful at all}.

\begin{figure}[htb]
\centering
\includegraphics[width=\columnwidth]{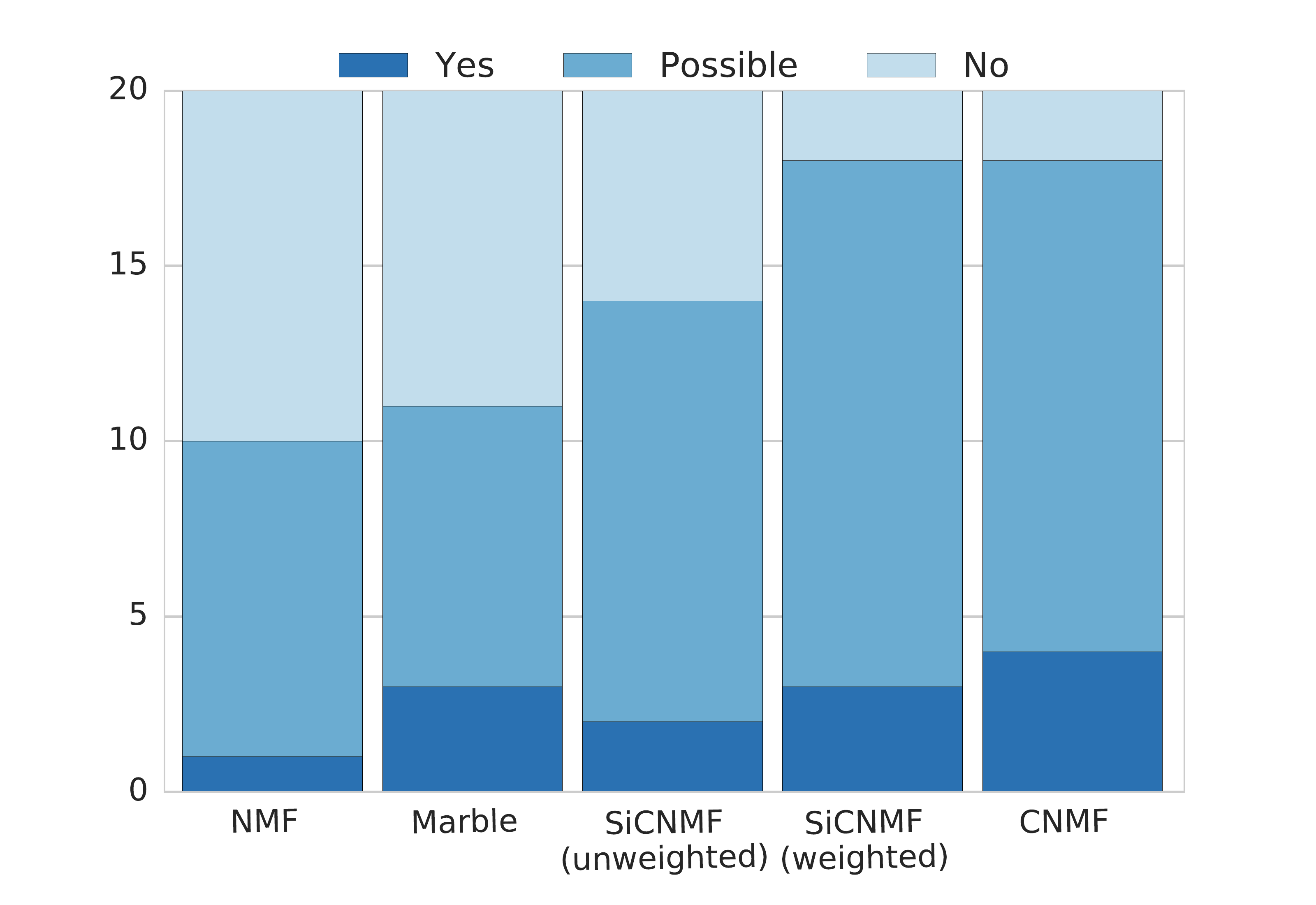}
\caption{The distribution of the clinical relevance scores across the various models.}
\label{fig:annotation-results}
\end{figure}

The annotated results for the models are compared  in Figure \ref{fig:annotation-results}.
The results show that weighted CMF based algorithms  perform significantly better in producing potentially clinical meaningful groupings. 
In an earlier work, Ho~et. al. \cite{Ho2014:KDD} show that for tensor valued data, Marble is very effective for  phenotyping. The improved performance of CMF based algorithms compared to Marble signifies the shortcomings of approximating the tensor from flat files, besides the additional computational cost of factorizing higher order tensors. 
Moreover, the improved performance of weighted SiCNMF compared to unweighted SiCNMF corroborates the efficacy of the weighing scheme described in Section~\ref{sec:alpha}. 

We note that although we could have performed SiCNMF on data that contains only the case patients to potentially yield more clinically relevant phenotypes, the purpose of our experiment was to demonstrate the unsupervised nature of our algorithm on a heterogeneous patient population.


Finally, Tables \ref{tab:sicnmf-example} and \ref{tab:cnmf-example} show examples of phenotypes derived from weighted SiCNMF and CNMF,  that were rated to be clinically meaningful by the domain experts.

\begin{table*}[htb]
\centering
\begin{tabular}{p{8cm} p{6cm}}
\toprule
 Diagnosis & Medication \\
\midrule
ischemic heart disease; hypertension; disorders of lipoid metabolism; late effects of cerebrovascular disease; occlusion of cerebral arteries; & antihyperlipidemic agents; cholinesterase inhibitors; antianginal agents; analgesics; antiplatelet agents; \\
\hline
chronic airway obstruction, not elsewhere classified; other diseases of lung; dyspnea and respiratory abnormalities; pneumonia, organism unspecified; hypertension; & bronchodilators; antiarrhythmic agents; calcium channel blocking agents; antiviral agents; medical gas;\\
\hline
malignant neoplasm of colon; rheumatoid arthritis and other inflammatory polyarthropathies; malignant neoplasm of rectum, rectosigmoid junction, and anus; secondary malignant neoplasm of respiratory and digestive systems; disorders involving the immune mechanism; & immunosuppressive agents; antirheumatics; antimetabolites; antipsoriatics; adrenal cortical steroid;\\
\bottomrule
\end{tabular}
\caption{Phenotypes from weighted--SiCNMF ($\eta=500$) that were evaluated as ``clinically meaningful" by our domain expert.
\label{tab:sicnmf-example}}
\end{table*}

\begin{table*}[htb]
\centering
\begin{tabular}{p{8cm} p{6cm}}
\toprule
 Diagnosis & Medication \\
\midrule
ischemic heart disease; hypertension; disorders of lipoid metabolism; unspecified chest pain; myocardial infarction; &
antianginal agents; antihyperlipidemic agents; vasodilators; antiplatelet agents; angiotensin converting enzyme inhibitors \\
\hline
heart failure; atrial fibrillation and flutter; hypertension; pulmonary heart disease; dyspnea and respiratory abnormalities; & diuretics;  antiarrhythmic agents; calcium channel blocking agents
bronchodilators;aldosterone receptor antagonists;
\\
\hline
malignant neoplasm of colon; rheumatoid arthritis and other inflammatory polyarthropathies; regional enteritis; malignant neoplasm of rectum, rectosigmoid junction, and anus; ulcerative colitis; & immunosuppressive agents; antirheumatics; analgesics; vitamins; antimetabolites;\\
\hline
chronic kidney disease (CKD); diabetes mellitus, type 2
Complications peculiar to certain specified procedures; other and unspecified anemias; diabetes mellitus, Type 1; & antidiabetic agents; miscellaneous antibiotics; sulfonamides; recombinant human erythropoietins; glucose elevating agents; \\
\bottomrule
\end{tabular}
\caption{Phenotypes from CNMF (no sparsity constraints) that were evaluated as clinically meaningful by our domain expert.
\label{tab:cnmf-example}}
\end{table*}

\section{Conclusions}
Unsupervised learning approaches for automated phenotyping have the potential to enable improved clinical trials, properly target patients for screening tests and interventions, and support surveillance of infectious diseases. However, traditional dimensionality reduction tools are not immediately well--suited for the phenotype extraction process. 
In this paper, we have introduced an unsupervised, structured collective matrix factorization tool that incorporates various application specific constraints into a joint low rank factorization framework. 
This framework is used for phenotype extraction from multi--source EHR data from Vanderbilt University. The clinical relevance  of extracted candidate phenotypes were evaluated by a domain expert and the results show improved performance over existing baseline models.

We intend to extend our studies along several directions. EHR data is often subject to noise and missing data. As a first extension of this work, we plan  to quantitatively study the robustness of the extracted phenotypes to (a) missing data, (b) noise in data, and (c) varying patient populations. Preliminary results in this direction are very encouraging. Secondly, in the current framework, we post--process the candidate phenotypes to enforce  hard sparsity requirements. We would like to explore models and algorithms that implicitly incorporate such sparsity constraints, potentially along the lines of the sparse simplex projection by \cite{kyrillidis2012sparse}. 

\bibliography{bibliography,healthcare,phenotyping_cnmf}
\end{document}